\documentclass{article}
\usepackage{ijcai24}

\usepackage{times}
\usepackage{soul}
\usepackage{url}
\usepackage[hidelinks]{hyperref}
\usepackage[utf8]{inputenc}
\usepackage[small]{caption}
\usepackage{graphicx}
\usepackage{amsmath}
\usepackage{amsthm}
\usepackage{booktabs}
\usepackage{algorithm}
\usepackage{algorithmic}
\usepackage{amsfonts}
\usepackage{subfig}
\usepackage{pgf-pie}
\usepackage[switch]{lineno}

\title{End-to-End Real-World Polyphonic Piano Audio-to-Score Transcription with Hierarchical Decoding}

\author{
Wei Zeng$^1$
\and
Xian He$^2$\And
Ye Wang$^{1,2}$
\affiliations
$^1$Integrative Sciences and Engineering Programme, NUS Graduate School\\
$^2$School of Computing, National University of Singapore\\
\emails
\{w.zeng, xian.he\}@u.nus.edu,
wangye@comp.nus.edu.sg
}

\begin{document}

\maketitle

\begin{abstract}
   Piano audio-to-score transcription (A2S) is an important yet underexplored task with extensive applications for music composition, practice, and analysis. However, existing end-to-end piano A2S systems faced difficulties in retrieving bar-level information such as key and time signatures, and have been trained and evaluated with only synthetic data. To address these limitations, we propose a sequence-to-sequence (Seq2Seq) model with a hierarchical decoder that aligns with the hierarchical structure of musical scores, enabling the transcription of score information at both the bar and note levels by multi-task learning. To bridge the gap between synthetic data and recordings of human performance, we propose a two-stage training scheme, which involves pre-training the model using an expressive performance rendering (EPR) system on synthetic audio, followed by fine-tuning the model using recordings of human performance. To preserve the voicing structure for score reconstruction, we propose a pre-processing method for **Kern scores in scenarios with an unconstrained number of voices. Experimental results support the effectiveness of our proposed approaches, in terms of both transcription performance on synthetic audio data in comparison to the current state-of-the-art, and the first experiment on human recordings.\footnote{Our code, pre-trained models, and demos can be found at \url{https://github.com/wei-zeng98/piano-a2s}.}
\end{abstract}

\section{Introduction}
Audio-to-score transcription (A2S) is a notation-level automatic music transcription (AMT) task that aims to transcribe audio into human- or machine-readable musical scores. By providing detailed notations  of musical compositions from audio recordings, A2S functions as a useful tool for music performance and music content analysis \cite{shibata2021non}. The piano, inherently polyphonic and covering a wide range of pitch, serves as an ideal model, offering potential extensions to other polyphonic music forms. A2S usually involves several interrelated subtasks, such as multi-pitch detection, rhythm quantization, voice separation, and key/time signature estimation.

Early A2S work \cite{cogliati2016transcribing,nakamura2018towards,shibata2021non} majorly decomposes A2S into subtasks and addresses them individually. From \cite{carvalho2017towards}, there has been a notable trend towards using end-to-end models for A2S, such as general Seq2Seq models \cite{liu2021joint} and connectionist temporal classification (CTC) models \cite{roman2018end,roman2019holistic,arroyo2022neural}. Compared to early decomposition-based methods, these end-to-end models tackle A2S more comprehensively, mitigating the potential problem of error accumulation from different sub-tasks.

However, the current state of end-to-end A2S models is still unfledged and thus cannot well address the following challenges.

\paragraph{Complexity of musical structures} Musical scores include various elements at different hierarchical levels \cite{koelsch2013processing}, such as notes, keys, and time signatures. This structural complexity is difficult to be well modelled by previous end-to-end A2S models, as existing studies \cite{liu2021joint,roman2019holistic,arroyo2022neural} primarily focus on note transcription, neglecting key and time signature transcription. Additionally, the polyphonic nature of piano music adds to the challenge, particularly in scenarios with an unconstrained number of voices.

\paragraph{Real-world evaluation gap} The availability of datasets for A2S transcription is limited, especially when compared to frame-level transcription datasets \cite{hawthorne2018enabling}. Existing end-to-end A2S systems primarily rely on synthetic audio directly from the quantized score MIDI for training and evaluation, which fails to capture human expressions in tempo, dynamics, and articulations. This reliance ignores the discrepancies between synthetic data and real-world recordings of human performance, limiting the real-world application of these models. To date, there has been no empirical evaluation of these end-to-end A2S systems using real-world audio recordings, underscoring a critical gap in the field.

\begin{figure*}[ht]
  \centering
  \includegraphics[width=1\linewidth]{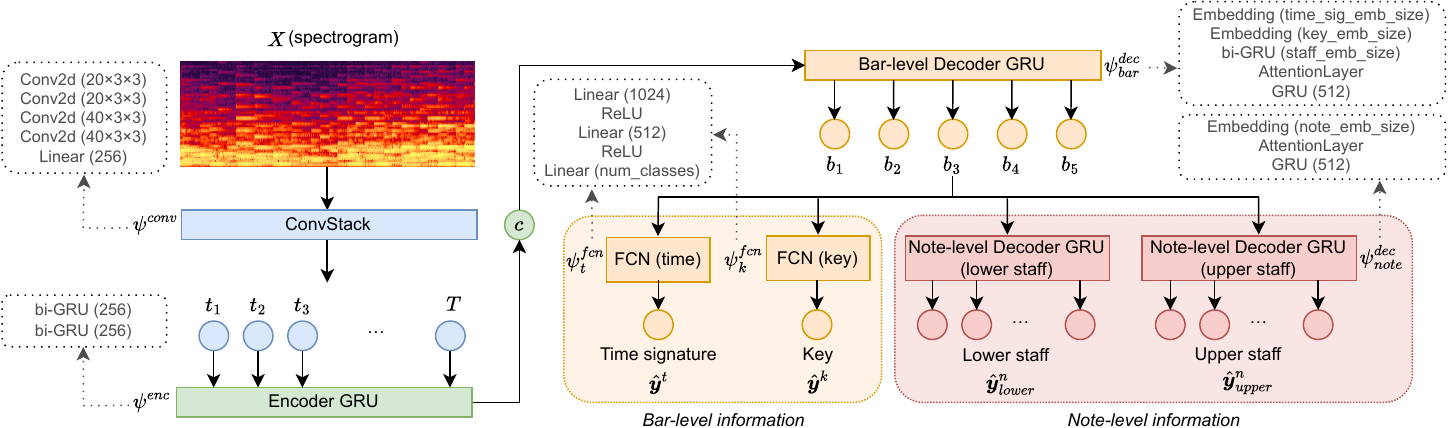}
  \caption{\label{fig:a2s}The proposed piano A2S model with a hierarchical decoder to transcribe the audio into both bar-level and note-level information. The model first encodes a Variable-Q Transform spectrogram $X$ into audio context representation $c$. Subsequently, a bar-level Decoder decodes $c$ into bar-level representation $b_i$ for the $i$-th bar. Time signatures ($\hat{\boldsymbol{y}}^t$) and keys ($\hat{\boldsymbol{y}}^k$) are transcribed at this bar-level given bar-level representations $\boldsymbol{b}$, while two note-level Decoders further decode $\boldsymbol{b}$ into note sequences of the lower staff ($\hat{\boldsymbol{y}}^n_{lower}$) and upper staff ($\hat{\boldsymbol{y}}^n_{upper}$), respectively.}
\end{figure*}

In this paper, we introduce a novel Seq2Seq A2S model, leveraging a hierarchical decoder to align with the hierarchical structure of piano music. As depicted in \autoref{fig:a2s}, the decoder first generates bar-level representations from the encoded audio representation, where key and time signatures are predicted. Subsequently, it further decodes these bar-level representations into note-level representations for note transcriptions. To bridge the gap between synthetic sound and real-world recordings of human performance, we propose a two-stage training scheme, which pre-trains the model on a large synthetic dataset generated from the state-of-the-art expressive performance rendering (EPR) system \cite{jeong2019virtuosonet}, and then fine-tunes the model on real-world piano recordings from human performance. Additionally, we extend previous work on score representation \cite{arroyo2022neural} by proposing a pre-processing method to serialize **Kern scores into token sequences while preserving the voicing structure. This representation strategy allows for the reconstruction of the final musical scores from predicted sequences. Our model is evaluated on both synthetic data and real piano recordings selected from the ASAP dataset \cite{foscarin2020asap}. These experiments demonstrate improvements over prior end-to-end systems and underscore the efficacy of our proposed training scheme.

To summarize, our main contributions are:
\begin{itemize}
    \item \textbf{A hierarchical A2S model by multi-task learning}: We introduce a novel end-to-end A2S model with a hierarchical decoder to transcribe the audio into both bar-level information, including key and time signatures, and note-level information, i.e., the note sequence.
    \item \textbf{A two-stage training scheme for A2S:} To bridge the gap between synthetic sound and real-world recordings, we first pre-train the model on a synthetic dataset generated from an EPR system and subsequently fine-tune it using real-world piano recordings from human performance. Experiments on human recordings show the effectiveness of our proposed training scheme.
    \item \textbf{A score representation method for unconstrained voices:} We introduce a pre-processing approach to serialize **Kern piano scores into tokens while preserving their inherent voicing structure, facilitating the reconstruction of scores with unconstrained number of voices.
    
\end{itemize}

\section{Related Work}

\subsection{A2S and Score Representation}
Existing end-to-end methods transcribe audio directly into score representations like LilyPond and **Kern. \cite{liu2021joint}, for example, used a general Seq2Seq model to predict target sequences for both left and right hands in the LilyPond format. Meanwhile, \cite{roman2019holistic} and \cite{arroyo2022neural} applied CTC loss to predict the target score sequence in the **Kern format. Among these studies, \cite{liu2021joint} and \cite{arroyo2022neural} emphasized polyphonic representation in their work. However, \cite{liu2021joint} is constrained to scenarios with only one voice per staff, thereby restricting its applicability to scores containing multiple voices occurring simultaneously for one hand. While \cite{arroyo2022neural} first attempted to tackle the unconstrained voices problem, their approach compromised the voicing structure during the serialization of the score, making it difficult to reconstruct the transcribed musical scores from the predictions.

\subsection{Expressive Performance Rendering Systems}
EPR systems are used for generating human-like performance MIDI given score files by delivering emotions and messages through subtle controls of tempo, dynamics, articulations and other expressive elements. Recent EPR systems have achieved promising results using neural network-based approaches \cite{jeong2019virtuosonet,jeong2019graph,renault2023expressive}. Among these systems, \cite{jeong2019virtuosonet} combined recurrent neural network (RNN) with hierarchical attention network and conditional variational autoencoder (CVAE), allowing users to specify a composer as an input for the output performance style.

\section{Methodology}

\subsection{Problem Formulation}
The primary objective of the A2S task is to transcribe provided piano recordings into sequences of musical notes containing pitch and note values, along with time and key signatures. Initially, the audio recordings are segmented into 5-bar clips, with segmentation based on the downbeat of each bar. These audio clips are then transformed into spectrograms, denoted as $X \in \mathbb{R}^{T \times F}$, where $T$ is the number of time frames and $F$ is the number of frequency bins.

Given the spectrograms as input, the A2S model predicts the note sequences in **Kern format for both the lower ($\boldsymbol{y}^n_{lower}$) and upper ($\boldsymbol{y}^n_{upper}$) staffs, as well as the sequences of time ($\boldsymbol{y}^t$) and key ($\boldsymbol{y}^k$) signatures of each bar respectively.

Specifically, the note sequence is denoted as $\boldsymbol{y}^n = (y^n_1, y^n_2, \cdots, y^n_{N})$, where each $y^n_i \in \Sigma ^{n'}$ and $N$ is the length of the sequence. Here, $\Sigma ^{n'}$ comprises the **Kern score representation vocabulary $\Sigma^n$ and special symbols $\{\langle \text{SOS} \rangle, \langle \text{EOS} \rangle, \langle \text{PAD} \rangle\}$. The time and key signatures are represented in sequences with a length of 5 corresponding to each bar: $\boldsymbol{y}^t = (y^t_1, y^t_2, \cdots, y^t_{5})$, and $\boldsymbol{y}^k = (y^k_1, y^k_2, \cdots, y^k_{5})$, where $y^t_i \in \Sigma ^{t}$ and $y^k_i \in \Sigma ^{k}$. $\Sigma ^{t}$ and $\Sigma ^{k}$ are vocabularies for time and key signatures, respectively. We keep 7 common time signatures for $\Sigma ^{t}$ as summarized in \autoref{tab:distribution}, and 14 keys for $\Sigma ^{k}$. The representation of $\Sigma ^{n}$ is detailed in Section \ref{sec:representation}.

\subsection{Data Representation}
\label{sec:representation}
\subsubsection{Introduction to **Kern Representation}
\begin{figure}
\centering
\includegraphics[width=0.9\linewidth]{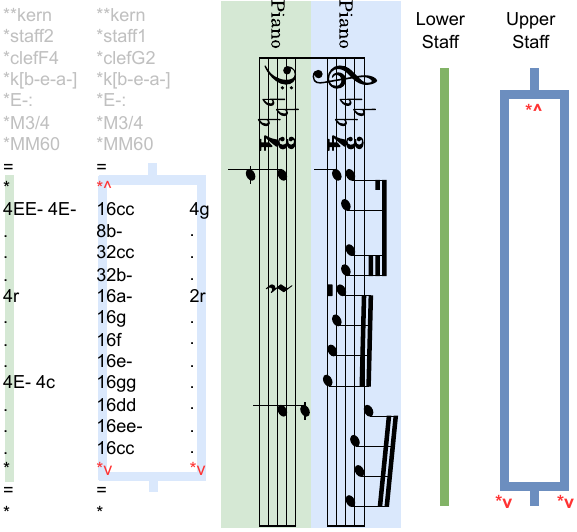}
\caption{\label{fig:kern_sample}A sample excerpt of multiple voicing: On the left, **Kern representation. On the center, the original score. On the right, the voicing structure of the excerpt. After pre-processing, the lower staff is serialized as: $\{4, EE-, \langle b \rangle, 4, E-, \backslash n, 4, r ...\}$, and the upper staff is serialized as :$\{16, cc, \backslash t, 4, g, \backslash n, 8, b-, ...\}$, where $\langle b \rangle$ is the token for a black space. The identifiers are manually added as post-processing.}
\end{figure}

We adopted **Kern as our score representation due to its clear structure, well-organized structure that accommodates multiple voices, and the abundance of data from the Humdrum database, from which we collect our HumSyn dataset as detailed in Section \ref{sec:data}. \autoref{fig:kern_sample} shows an example of a **Kern representation for a one-bar piano score together with its original musical score and simplified voicing structure. In this example, the lower staff consists of a single voice, while the upper staff consists of two voices. Consequently, the **Kern score is encoded with three \textit{spines}, each corresponding to a distinct voice. It is important to distinguish between polyphonic and multiple voices: in the lower staff of the excerpt, there are two notes occurring simultaneously, indicating its polyphonic characteristic, but these simultaneous notes belong to a single voice. In a scenario with unconstrained voices, multiple voices may appear within the same staff, as shown in the upper staff of the excerpt. To manage this, **Kern representation allows for the division and merging of spines using specific identifiers, namely `*\^{}' and `*v', which aids in maintaining and tracking the voicing structure. In **Kern representation, concurrent notes within a voice are separated by blank spaces, while notes belonging to different voices are separated by tabs.

The voicing structure within **Kern representation heavily relies on accurate predictions of the identifiers. An incorrect prediction of these identifiers can result in the loss of voicing structure, posing significant challenges during the reconstruction process. For instance, as illustrated in the voicing structure of \autoref{fig:kern_sample}, if the model fails to make a correct prediction regarding the `*\^{}' token, the middle voice is ambiguous to either the lower staff or the upper staff, misleading the reconstruction process.

\subsubsection{Proposed Pre-processing Method}
To gain more insights on the voicing structure in piano scores for better pre-processing, we analyzed 235 scores from the ASAP dataset, and counted the number of voices present in each staff at the bar level. The results revealed that bars containing no more than two voices per staff comprised 95.2\% of the total in the lower staff and 97.6\% in the upper staff. Taking into account the possibility of notation errors in the score files, this observation leads us to a robust conclusion: in the majority of scenarios, a single staff typically contains no more than two voices. Building upon this premise, we propose to separate the **Kern representation into lower and upper staffs and predict them individually within the model. This approach allows us to eliminate the identifiers within each staff without introducing ambiguity to the voicing structure. To reconstruct the score from the prediction, we manually reintroduce the identifiers during the post-processing stage, specifically before each transition between one voice and two voices.

Additionally, **Kern does not constrain the order of notes within chords or between different voices. It also allows for the inclusion of empty voices within scores. To ensure consistency in our data representation, we employ the following steps as pre-processing:
\begin{itemize}
    \item Separate the lower staff and upper staff individually, and remove the identifiers `*\^{}' and `*v' from the **Kern representation for each staff.
    \item Eliminate empty voices, and merge two parallel voices if their notes have identical durations.
    \item Arrange concurrent notes within the same voice in ascending order of pitch.
    \item Organize the voices so that the higher voice appears on the left stem, while the lower voice appears on the right stem.
\end{itemize}

\subsection{Model Architecture}

The A2S model is comprised of a ConvStack ($\psi^{conv}$), an Encoder ($\psi^{enc}$), a bar-level Decoder ($\psi^{dec}_{bar}$), note-level Decoders ($\psi^{dec}_{note}$) and Fully Connected Network (FCN) layers ($\psi^{fcn}_t$, $\psi^{fcn}_k$), as shown in \autoref{fig:a2s}. The input spectrogram firstly passes through $\psi^{conv}$ for feature extraction. We use padding to fix the time dimension, such that

\begin{equation}
    \psi^{conv}(X) \in \mathbb{R} ^{T \times d},
\end{equation}
where $d$ is the feature dimension of the output of ConvStack.

Next, the Encoder encodes the features into a single representation:
\begin{equation}
    outs, c = \psi^{enc} (\psi^{conv}(X)),
\end{equation}
where $outs \in \mathbb{R} ^{T \times d}$ is the outputs of the last layers, and $c \in \mathbb{R} ^{d}$ is the audio content representation taken from the last hidden state.

For the Decoder, we developed a hierarchical structure capable of handling both bar-level and note-level information. Starting from the Encoder outputs $outs$ and the encoded audio representation $c$, we employ the bar-level Decoder $\psi^{dec}_{bar}$ to decode it into bar representations, where each $b_i$ denotes the feature representation for the $i$-th bar. At the bar level, FCN layers $\psi^{fcn}_t$ and $\psi^{fcn}_k$ are employed to make predictions for the time signature $\hat{y}_i^t$ and key signature $\hat{y}_i^k$ given $b_i$. Subsequently, using these bar-level representations as inputs, two note-level Decoders decode the music notes into **Kern sequences for the lower and the upper staff respectively, resulting in the predicted note sequences $\hat{\boldsymbol{y}}^n_{lower}$ and $\hat{\boldsymbol{y}}^n_{upper}$. This hierarchical decoding process is summarized in Algorithm \ref{alg:decoder} in more details.

\begin{algorithm}[tb]
            \caption{The Hierarchical Decoder}
    \label{alg:decoder}
    \textbf{Input}: Encoded audio content representation $c$, Encoder outputs of the last layer $outs$ \\
    \textbf{Parameter}: Bar-level hidden state $\boldsymbol{h}^b$, bar-level token $\widetilde{\boldsymbol{b}}$, bar-level feature $\boldsymbol{b}$, bar-level attention $\boldsymbol{a}$, note-level hidden state $\boldsymbol{h}^n$, note-level token $\widetilde{\boldsymbol{n}}$, note-level feature $\boldsymbol{n}$, attention layer Attn($\cdot$), embedding layer Emb($\cdot$) \\
    \textbf{Output}: Time signature $\hat{\boldsymbol{y}}^t$, key signature $\hat{\boldsymbol{y}}^k$, and note sequences $\hat{\boldsymbol{y}}^n$ ($\hat{\boldsymbol{y}}^n_{lower}$, $\hat{\boldsymbol{y}}^n_{upper}$)
    \begin{algorithmic}[1] 
        \STATE Let $h^b_0=c$, $\widetilde{b}_0 = \langle \text{SOS} \rangle $.
        \FOR{i = 1, 2, \dots, 5}
        \STATE $a_i = \text{Attn} (h^b_{i-1}, outs)$
        \STATE $b_i, h^b_i = \psi^{dec}_{bar} (\text{Cat}(\widetilde{b}_{i-1}, a_i), h^b_{i-1})$
        \STATE $\hat{y}^t_i = \text{log}\_\text{softmax}(\psi^{fcn}_t(\text{Cat}(b_i, a_i)))$
        \STATE $\hat{y}^k_i = \text{log}\_\text{softmax}(\psi^{fcn}_k(\text{Cat}(b_i, a_i)))$
        \STATE Let $h^n_{i,0} = b_i$, $\widetilde{n}_{i,0} =\langle \text{SOS} \rangle$
        \FOR{t = 1, 2, \dots, max\_steps}
        \STATE $n_{i,t}, h^n_{i,t} = \psi^{dec}_{note} (\text{Cat} (\widetilde{n}_{i,t-1}, \text{Attn} (h^n_{i,t-1}, outs))$, \\
        \qquad \qquad \qquad \quad $h^n_{i,t-1})$
        \STATE $\hat{y}^n_{i,t} = \text{log}\_\text{softmax}(n_{i,t})$
        \STATE $\widetilde{n}_{i,t} = \text{Emb}(\text{argmax}(\hat{y}^n_{i,t}))$
        \IF{$\widetilde{n}_{i,t} == \langle \text{EOS} \rangle$}
        \STATE break
        \ENDIF
        \ENDFOR
        \STATE staff\_emb = GRU ($\hat{y}^n_{i, 1:\text{max}\_\text{steps}}$)
        \STATE $\widetilde{b}_i = \text{Cat}(\text{staff}\_\text{emb}, \text{Emb}(\text{argmax}(\hat{y}^t_i))$, \\
        \qquad \qquad $\text{Emb}(\text{argmax}(\hat{y}^k_i)))$
        \ENDFOR
        \STATE \textbf{return} $\hat{\boldsymbol{y}}^t$, $\hat{\boldsymbol{y}}^k$, $\hat{\boldsymbol{y}}^n$
    \end{algorithmic}
\end{algorithm}

For the training, we employ multi-task learning \cite{zhang2021survey}. We use the negative log-likelihood loss (NLLL) for each subtask, which includes key, time, and note sequences for both the lower and upper staffs:
\begin{equation}
l_{\text{subtask}} = \text{NLLL}(\hat{\boldsymbol{y}}^{\text{subtask}}, \boldsymbol{y}^{\text{subtask}})
\end{equation}
The total loss is the summation of these individual losses:
\begin{align}
l &= l_{\text{key}} + l_{\text{time}} + l_{\text{lower}} + l_{\text{upper}}.
\end{align}

\subsection{Training Scheme}

Synthetic data has been extensively used in A2S tasks due to its ease of access and potential for augmentation. In contrast, obtaining real piano recordings from human performance is challenging and less accessible. In order to bridge the gap between synthetic sounds and real human performance while leveraging the easy accessibility of synthetic sounds, we propose a two-stage training scheme, as illustrated in \autoref{fig:scheme}. The pre-training leverages data synthesized from the state-of-the-art EPR model, VirtuosoNet \cite{jeong2019virtuosonet}. Compared to data directly generated from score MIDI, these synthetic data are more similar to human performance as they capture some subtle details in piano performance such as deviations in note onsets, durations, velocities, and pedal usage. The model is pre-trained on synthetic audio from the EPR system. In the subsequent fine-tuning stage, we handpick a small set of real-world recordings of human performance from the ASAP dataset \cite{foscarin2020asap} to fine-tune the pre-trained model by transfer learning.

\section{Experiments}
\subsection{Data Preparing}
\label{sec:data}
\begin{figure*}[ht]
  \centering
  \includegraphics[width=1\linewidth]{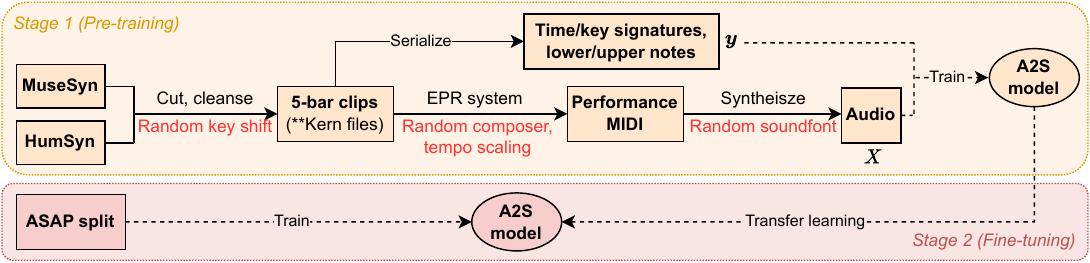}
  \caption{\label{fig:scheme}The two-stage training scheme including pre-training on synthetic data from an EPR system, and fine-tuning on human recordings.}
\end{figure*}

\subsubsection{Corpora} 
For the pre-training stage, we consider two score corpora: MuseScore website\footnote{\url{https://musescore.com/hub/piano}} and Humdrum repository\footnote{\url{http://kern.ccarh.org/}}. From MuseScore, we choose the same MuseSyn dataset used in \cite{liu2021joint}, primarily comprising pop music. From the Humdrum repository, we select piano sonatas by Beethoven\footnote{\url{https://github.com/craigsapp/beethoven-piano-sonatas.git}}, Haydn\footnote{\url{https://github.com/craigsapp/haydn-piano-sonatas.git}}, Mozart\footnote{\url{https://github.com/craigsapp/mozart-piano-sonatas.git}}, Scarlatti\footnote{\url{https://github.com/craigsapp/scarlatti-keyboard-sonatas.git}}, as well Joplin's compositions\footnote{\url{https://github.com/craigsapp/joplin.git}} and Chopin's works\footnote{\url{https://github.com/pl-wnifc/humdrum-chopin-first-editions.git}}, covering a range of styles from classical to ragtime. This combined corpus from Humdrum database is referred to as the HumSyn dataset.

For pre-training, we randomly select 80\% of the pieces from the HumSyn dataset and combine them with the training split of the MuseSyn dataset together as the training set. Similarly, we create validation and test sets by randomly selecting 10\% of the pieces from the HumSyn dataset and combining them with the corresponding splits of the MuseSyn dataset. This is to ensure a consistency on the MuseSyn test split with \cite{liu2021joint} for a fair comparison. 

For the fine-tuning, we use the ASAP dataset \cite{foscarin2020asap} with annotated alignments between scores and human performances. From ASAP, we select 14 songs, each featuring recordings by one or more different performers, with a total of 58 recordings in our chosen subset. Additionally, we choose 25 different compositions for the test set, including a total of 80 recordings performed by various artists. We confirmed that these splits are disjoint and cannot lead to data leakages. The statistics about MuseSyn, HumSyn, and ASAP split are summarized in \autoref{tab:distribution}.

\begin{table}
    \centering
    \begin{tabular}{lrrrr}
        \toprule
        Dataset  & Split & Num & Time Signatures \\
        \midrule
        \textbf{MuseSyn}   & train & 169 & 4/4,3/4,2/4,6/8,2/2,12/8        \\
        \textit{(Pop)}          & valid & 20 & 4/4,3/4,2/2         \\
                  & test  & 21 & 4/4,3/4,6/8        \\
        \midrule
        \textbf{HumSyn}    & train & 382 & 4/4,3/4,2/4,6/8,2/2,12/8,3/8       \\
        \textit{(Classical,}      & valid & 46 & 4/4,3/4,2/4,6/8,2/2,3/8         \\
        \textit{Ragtime)}      & test  & 45 & 4/4,3/4,2/4,6/8,2/2,12/8,3/8 \\
        \midrule
        \textbf{ASAP}      & train & 14 & 4/4,3/4,2/4,2/2 \\
        \textit{(Classical)}          & test  & 25 & 4/4,3/4,2/4,6/8,2/2,3/8       \\
        \bottomrule
    \end{tabular}
    \caption{Statistics about number of musical scores and time signatures in different splits used in the experiments.}
    \label{tab:distribution}
\end{table}

\subsubsection{Data Synthesis Pipeline}

The data synthesis pipeline for the pre-training stage is illustrated in \autoref{fig:scheme}. Firstly, we segment the scores into 5-bar clips and cleanse the scores to eliminate ornamental elements such as editorial marks and grace notes. 

For the training process, we employ four sequential steps for data augmentation on each of the segmented clip, including random key shifts within a range of four semitones, random selection from 15 composers within the EPR system \cite{jeong2019virtuosonet}, random tempo scaling between 0.85 and 1.15 for the performance MIDI, and the random selection from four piano soundfonts during the synthesizing process. These data augmentation methods expand the final training set to 10$\times$ of its original size, with key signatures spanning a wide range covering number of sharps from -6 to 7.

For the validation and testing processes, we keep the original key and tempo, while generating performance MIDI using four EPR composers for every score clip: \textit{Score (}generating MIDI directly from the score, \textit{id)}, \textit{Bach (id)}, \textit{Mozart (ood)}, and \textit{Chopin (ood)}. The term \textit{id}, or in-distribution, indicates the composer is used in training, while \textit{ood}, or out-of-distribution, is not. Additionally, we synthesize audio using three distinct piano soundfonts for every MIDI clip: \textit{UprightPianoKW (Upright, id)}, \textit{SalamanderGrandPiano (Salamander, id)} and \textit{YDP-GrandPiano (YDP, ood)}.

\subsection{Metrics}
We evaluate the score transcription using two established metrics, in accordance with prior end-to-end A2S research \cite{liu2021joint,arroyo2022neural,roman2019holistic}: word error rate (WER) and the MV2H metric \cite{mcleod2018evaluating,mcleod2019evaluating}. Additionally, to evaluate the performance of key and time signature transcription, we employ the F1-score for bar-level transcriptions. Below are the details of these metrics:

\begin{itemize}
    \item \textbf{WER}: We calculate the WER by comparing the transcription score with the original in **Kern sequences, treating each individual token as a word. We report both the overall WER and the individual WER for the two staffs, denoted as $\text{WER}_\text{l}$ and $\text{WER}_\text{u}$ for the lower and upper staff, respectively.
    \item \textbf{MV2H}: MV2H evaluation includes four sub-metrics: multi-pitch detection accuracy ($F_p$), voice separation accuracy ($F_{voi}$), note value detection accuracy ($F_{val}$), and harmonic detection ($F_{harm}$). Since we do not constrain metrical prediction at the note level, we have excluded the sub-metric for metrical alignment. Instead, we predict the time signature at the bar level and assess its performance using the F1-score. The overall metric ($F_{MV2H}$) is computed as the average of the four sub-metrics. We employ the non-aligned version of MV2H \cite{mcleod2019evaluating}, which is suitable for our non-aligned scenario and allows for automatic alignment.
    \item \textbf{F1-score}: Due to the imbalanced distribution of key and time signature labels, we employ the macro F1-score for key ($f_{k}$) and time signature ($f_{t}$) respectively.
\end{itemize}

\subsection{Experimental Setup}
We build our hierarchical piano A2S model using the PyTorch library and SpeechBrain toolkit \cite{ravanelli2021speechbrain}. We transform the audio into variable-Q transform (VQT) spectrograms \cite{schorkhuber2014matlab}, following \cite{liu2021joint}. The audio sample rate and hop length are 16kHz and 160. We employ 60 bins per octave across a total of 8 octaves, and the gamma value for the VQT is 20. During the training process of the pre-training stage, we employ teacher forcing \cite{lamb2016professor} to expedite the model convergence. Initially, the teacher forcing ratio is set to 0.7, with a decay rate of 0.99 for each epoch. In the training process of the fine-tuning stage, the teacher forcing ratio remains fixed at 0.6, given the limited size of the dataset and knowledge acquired during the pre-training stage. 

\subsection{Comparison with Baseline Model}
\begin{table}
    \centering
    \begin{tabular}{lrrrrr}
        \toprule
        Model & $F_p$  & $F_{voi}$ & $F_{val}$ & $F_{harm}$ & $F_{MV2H}$  \\
        \midrule
        baseline (\textit{Joint}) & 71.1 & \textbf{90.8} & 94.4 & - & 85.4 \\
        \textit{Ours (score)} & \textbf{81.2} & 90.4 & \textbf{95.3} & 82.1 & \textbf{87.2} \\
        \textit{Ours (EPR)} & 80.0 & 87.7 & 92.4 & 84.0 & 86.0 \\
        \bottomrule
    \end{tabular}
    \caption{Comparison of our models with the state-of-the-art baseline model on MuseSyn test split synthesized from score MIDI.}
    \label{tab:joint}
\end{table}

We select the Joint model with Reshaped representation (\textit{Joint}) from \cite{liu2021joint} as our baseline model. This choice is motivated by several factors: it specializes in polyphonic piano transcription, adopts an end-to-end approach, provides a publicly available test split, and uses a comparable MV2H metric. In contrast, some other models, like \cite{arroyo2022neural}, randomize test split and are not directly comparable to our work due to the use of different tokenizers and reliance solely on the WER metric. 

We synthesize the same training split in two different ways: 1) synthesizing directly from the score MIDI, and 2) using the EPR system for synthesizing. Models pre-trained on these two training splits are denoted as \textit{Ours (score)} and \textit{Ours (EPR)} respectively. To enable a fair comparison, we synthesize the MuseSyn test split directly from score MIDI, following the methodology employed in \cite{liu2021joint}.

We compare our models with the baseline \textit{Joint} model on MuseSyn test split in terms of MV2H metrics. From \autoref{tab:joint}, it is evident that our model exhibits improvements over the baseline \textit{Joint} model, particularly in multi-pitch detection and note value detection. Notably, on multi-pitch detection,  \textit{Ours (score)} shows an increase of 10.1\%. Furthermore, we observe that \textit{Ours (score)} exhibits higher accuracy compared to \textit{Ours (EPR)} on the test split of MuseSyn synthesized from the score MIDI, but \textit{Ours (score)} and \textit{Ours (EPR)} perform the opposite on human recordings, as elaborated in the following sections.

\begin{table*}[ht]
    \centering
    \begin{tabular}{lrrrrrrrrrrr}
        \toprule
        Model & Fine-tune & $F_p$  & $F_{voi}$ & $F_{val}$ & $F_{harm}$ & $F_{MV2H}$ & $\text{WER}_\text{l}$ & $\text{WER}_\text{u}$ & WER & $f_{k}$ & $f_{t}$  \\
        \midrule
        \textit{Ours (score)} & No & 39.8 & 75.8 & 80.1 & 46.9 & 60.6 & 84.0 & 91.7 & 87.8 & \textbf{37.6} & 38.7 \\
        & Yes & 60.4 & 87.8 & \textbf{90.9} & 50.0 & 72.3 & 61.5 & 56.8 & 59.2 & 36.3 & 62.6 \\
        \midrule
        \textit{Ours (EPR)} & No & 53.5 & 83.2 & 89.8 & 51.4 & 69.4 & 67.0 & 65.7 & 66.4 & 36.6 & 56.4 \\
        & Yes & \textbf{63.3} & \textbf{88.4} & 90.7 & \textbf{54.5} & \textbf{74.2} & \textbf{57.0} & \textbf{53.2} & \textbf{55.1} & 36.5 & \textbf{71.2} \\
        \bottomrule
    \end{tabular}
    \caption{The MV2H, WER and F1-score results of \textit{Ours (score)} and \textit{Ours (EPR)} on ASAP test split before and after fine-tuning.}
    \label{tab:fine-tune}
\end{table*}

\begin{table}
    \centering
    \begin{tabular}{lrrrrr}
        \toprule
        Test Split & $F_p$  & $F_{voi}$ & $F_{val}$ & $F_{harm}$ & $F_{MV2H}$  \\
        \midrule
        \textit{Score (id)} & 79.8 & 87.3 & \textbf{92.8} & 65.9 & 81.4 \\
        \textit{Bach (id)} & \textbf{82.8} & 89.0 & 92.5 & 67.1 & 82.8 \\
        \textit{Mozart (ood)} & 82.0 & \textbf{89.2} & \textbf{92.8} & \textbf{67.5} & \textbf{82.9} \\
        \textit{Chopin (ood)} & 80.2 & 88.8 & 92.4 & 67.0 & 82.1 \\
        \midrule
        \textit{Upright (id)} & \textbf{83.5} & 89.5 & \textbf{92.9} & 67.0 & 83.2 \\
        \textit{Salamander (id)} & 83.2 & \textbf{90.0} & 92.8 & \textbf{67.1} & \textbf{83.3} \\
        \textit{YDP (ood)} & 76.9 & 86.3 & 92.1 & 66.5 & 80.4 \\
        \midrule
        \textit{Overall} & 81.2 & 88.6 & 92.6 & 66.9 & 82.3 \\
        \bottomrule
    \end{tabular}
    \caption{MV2H metric of \textit{Ours (EPR)} on test splits synthesized by different EPR composers and piano soundfonts.}
    \label{tab:pre-train_mv2h}
\end{table}

\subsection{Results from the Pre-training Stage}
\begin{figure}[ht]
  \centering
  \includegraphics[width=0.86\linewidth]{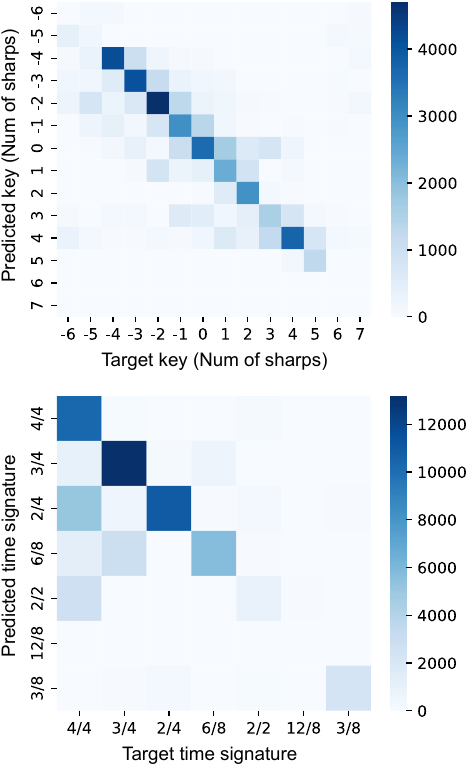}
  \caption{\label{fig:conf}The confusion matrices of key (upper) and time signature (lower) of \textit{Ours (EPR)} in the pre-training stage.}
\end{figure}

\subsubsection{Generalization Capability}
To further investigate the impact of EPR composers and piano soundfonts on \textit{Ours (EPR)}'s performance, as well as its ability to generalize to previously unseen performance styles and soundfonts in the pre-training stage, we summarized the performance of \textit{Ours (EPR)} on MV2H with respect to different composers and soundfonts in \autoref{tab:pre-train_mv2h}. The results reveal that the model demonstrates strong generalization capabilities to unseen performance styles. However, it is worth noting that \textit{ood} soundfonts can have a more significant influence on performance, with an acceptable degradation over \textit{id} soundfonts. This can be attributed to two main factors: firstly, the diversity in performance styles arising not only from EPR composers but also from various compositions themselves, enhancing the model's generalizability to unseen performance styles; and secondly, the model's exposure to a limited set of soundfonts during pre-training, resulting in a similarity of learned sound characteristics.

\subsubsection{Bar-level Information Transcription}
To gain further insights into the model's key and time signature prediction, we visualize the confusion matrices for the test split of \textit{Ours (EPR)}, as depicted in \autoref{fig:conf}. The matrices illustrate the number of bars along with their corresponding targets and predictions. In music theory, key distance is used to describe the varying degrees of relevance between musical keys. It is defined as the difference in the number of sharps or flats in the key signatures \cite{krumhansl1982key}. Therefore, in \autoref{fig:conf}, adjacent keys have the closest key distance. From \autoref{fig:conf}, we observe a pronounced trend in the key confusion matrix: the model tends to make mistakes in keys with close distances. Similarly, in time signature detection, the model shows a tendency to confuse similar time signatures, such as 4/4 and 2/2, 3/4 and 6/8, etc.

\begin{figure}[ht]
  \centering
  \includegraphics[width=1\linewidth]{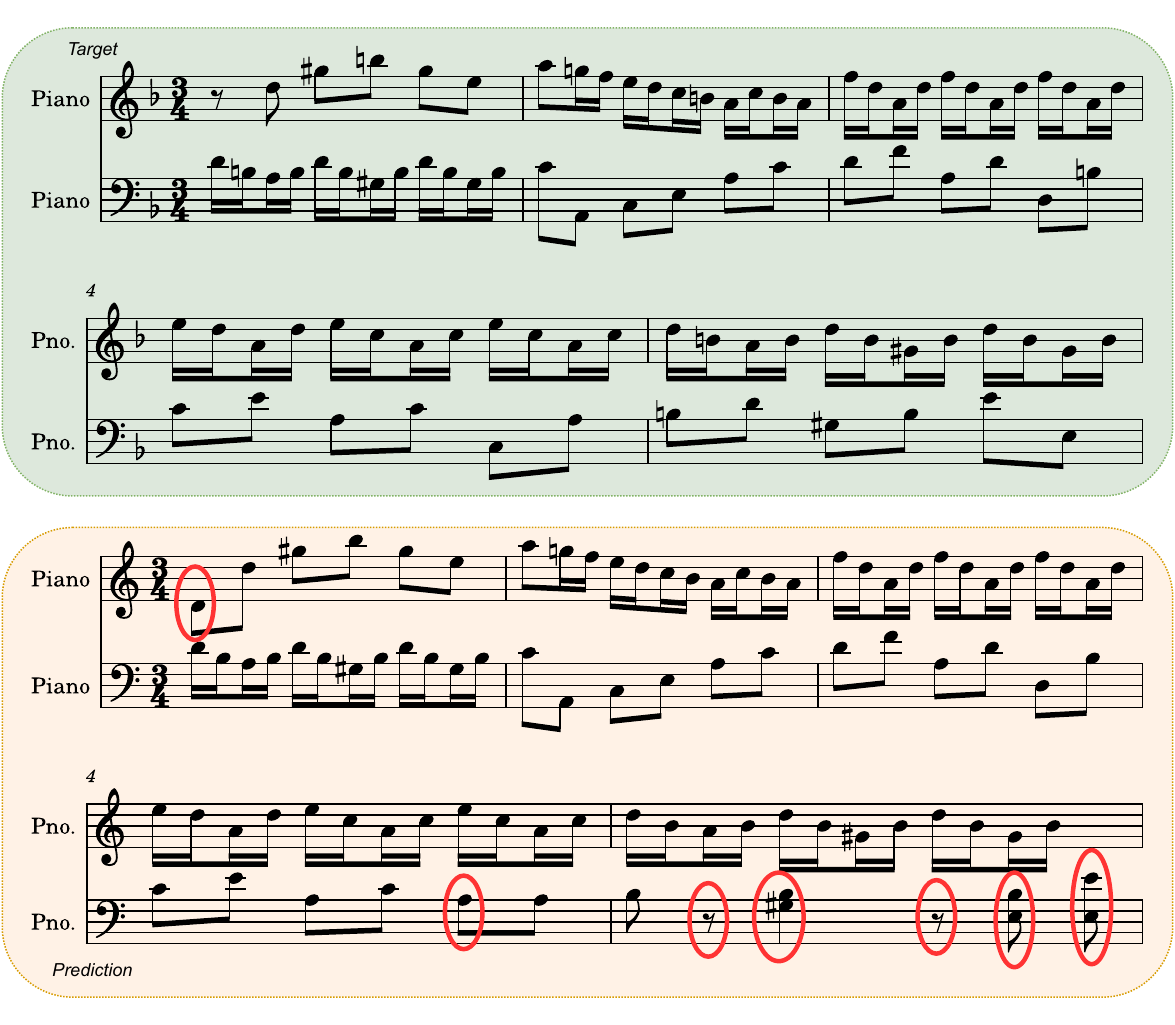}
  \caption{\label{fig:sample}A sample score (upper) and its transcription result (lower) from the fine-tuned \textit{Ours (EPR)} model. The excerpt is selected from \textit{Prelude and Fugue in D minor, BWV 875 (Bach, Johann Sebastian)}, performed by HONG04M.}
\end{figure}

\subsection{Results from the Fine-tuning Stage}
\subsubsection{Effectiveness of Using EPR Models for Pre-training}
To investigate the impact of the EPR system used in the pre-training stage, we compare the performance of \textit{Ours (score)} and \textit{Ours (EPR)} on the ASAP test split before and after fine-tuning. The MV2H, WER and F1-score are summarized in \autoref{tab:fine-tune}. From the table, it is evident that \textit{Ours (EPR)} outperforms \textit{Ours (score)} on almost all metrics, both before and after fine-tuning. Even with only a small number of songs selected for fine-tuning, both models exhibit improvements across most metrics. Taking these points above, we conclude that audio synthesized from EPR systems capture more similarities from human performances, which is a more efficient way for pre-training. These results underscore the effectiveness of utilizing EPR systems for pre-training.

\subsubsection{Case Study}
We present a sample result from the test ASAP split transcribed by \textit{Ours (EPR)} after fine-tuning on the ASAP training set, as shown in \autoref{fig:sample}. Notably, it is selected from real-world human recordings, with the model transcribing all score information. Upon closer examination, several interesting observations can be made. While the entire composition is based on the key of \textit{D} minor, as indicated by one flat in the upper part of \autoref{fig:sample}, there is a temporary shift to \textit{A} minor key within the audio clip. Remarkably, our model correctly identifies this potential key change, even though the key signature of the original piece remains the same. This demonstrates the model's capability to detect such key variations.

\section{Conclusion}

In this paper, we introduced a novel end-to-end piano A2S model targeting at two main challenges in existing work: difficulty in modeling hierarchical musical structures, and the discrepancies between synthetic data and real-world recordings from human performance. We proposed a Seq2Seq model with a hierarchical decoder that transcribes both the bar-level and note-level information. To bridge the gap between synthetic data and real-world recordings, we proposed a two-stage training scheme to pre-train the model on synthetic data from an EPR system, and fine-tune the model on human performance. Furthermore, to preserve the voicing structure in the musical score, we proposed a pre-processing method for **Kern representation for score reconstruction. We conducted experiments on the pre-trained model to evaluate its generalization capabilities in both in-distribution and out-of-distribution scenarios. We also conducted the first experiment for end-to-end A2S systems on piano recordings of human performance on ASAP dataset and validated the effectiveness of our proposed training scheme.

\section*{Acknowledgments}

We would like to thank anonymous reviewers for their valuable suggestions. We also appreciated Joon Siong Yap’s work on audio synthesis and Qihao Liang's writing advice.

\bibliographystyle{named}
\bibliography{ijcai24}

\end{document}